\documentclass[conference]{IEEEtran}

\usepackage{amsmath}
\usepackage[tight]{subfigure}
\usepackage[footnotesize]{caption} 
\usepackage{comment}
\usepackage{cite}
\usepackage{eurosym}
\usepackage{amsfonts}
\usepackage{graphics}
\usepackage{epsfig}
\usepackage{makecell}
\usepackage{psfrag}
\usepackage{pstricks}
\usepackage{array}
\usepackage{shortvrb}
\usepackage{epsf}
\usepackage{graphicx}
\usepackage{rotating}
\usepackage{float}
\usepackage{color}
\usepackage{cite}
\usepackage{soul}
\usepackage{multirow}
\usepackage{acronym}
\usepackage{mathtools} 
\usepackage{url}
\usepackage{breakurl}
\usepackage[breaklinks, hidelinks]{hyperref}

\acrodef{ess}[\textsc{ess}]{Energy Storage System}
\acrodef{dae}[\textsc{dae}]{Differential Algebraic Equation}
\acrodef{vsc}[\textsc{vsc}]{Voltage Sourced Converter}
\acrodef{coi}[\textsc{coi}]{Centre of Inertia}
\acrodef{fdf}[\textsc{fdf}]{Frequency Divider Formula}
\acrodef{wecs}[\textsc{wecs}]{Wind Energy Conversion System}
\acrodef{spvg}[\textsc{spvg}]{Solar Photo-Voltaic Generation}
\acrodef{tcl}[\textsc{tcl}]{Thermostatically Controlled Load}
\acrodef{hvdc}[\textsc{hvdc}]{High-Voltage Direct Current}
\acrodef{pll}[\textsc{pll}]{Phase-Locked Loop}
\acrodef{pmu}[\textsc{pmu}]{Phasor Measurement Unit}
\acrodef{rtds}[\textsc{rtds}]{Real-Time Digital Simulator}
\acrodef{emt}[\textsc{emt}]{Electromagnetic Transients}
\acrodef{tg}[\textsc{tg}]{Turbine Governor}
\acrodef{avr}[\textsc{avr}]{Automatic Voltage Regulator}
\acrodef{agc}[\textsc{agc}]{Automatic Generation Control}
\acrodef{gps}[\textsc{gps}]{Global Positioning Satellite}
\acrodef{rocof}[RoCoF]{Rate of Change of Frequency}

\graphicspath{{./pdf/}}
\DeclareGraphicsExtensions{.eps,.png,.pdf}

\def \R {{\rm I\kern -2.2pt R\hskip 1pt}}

\newcommand{\PreserveBackslash}[1]{\let\temp=\\#1\let\\=\temp}

\IEEEoverridecommandlockouts

\begin{document}

\title{Rethinking Frequency Control in Power Systems}

\author{Taulant K{\"e}r{\c c}i,~\IEEEmembership{Senior Member,~IEEE}, {\'A}ngel Vaca,~\IEEEmembership{Senior Member,~IEEE}, Andrew Groom, \\ Julia~Matevosyan,~\IEEEmembership{Fellow,~IEEE}, 
  and Federico~Milano,~\IEEEmembership{Fellow,~IEEE}
  \thanks{T.~K{\"e}r{\c c}i and {\'A}.~Vaca are with the Irish Transmission System Operator,
     EirGrid, Ireland.}
   \thanks{A.~Groom is with the Australian Energy Market Operator, Melbourne, Victoria, 3000, Australia.} 
   \thanks{J.~Matevosyan is with the Energy Systems Integration Group, Reston, VA 20195 USA.}
   \thanks{F.~Milano is with School of Electrical and Electronic Engineering,
    University College Dublin, Belfield Campus, Dublin 4, Ireland.
    Corresponding author.  e-mail: federico.milano@ucd.ie.}
  \thanks{This work was partially supported by Sustainable Energy
    Authority of Ireland (SEAI) by funding Federico Milano through
    FRESLIPS project, Grant No.~RDD/00681.}
  \vspace{-3mm}
  }

\maketitle

\begin{abstract}
  Frequency control in power systems is implemented in a hierarchical structure traditionally known as primary frequency control (PFC), secondary frequency control (SFC) and tertiary control reserve (TCR) and, some jurisdictions, include time error control (TEC) as well.  This hierarchical structure has been designed around a century ago based on timescales separation, that is, approximately an order of magnitude difference between each control structure.  This paper argues, based on real-world observations as well as detailed dynamic simulations on a model of the All-Island power system (AIPS) of Ireland, that this frequency control structure is not necessary in current and future converter-dominated power grids.  The paper proposes to redesign this structure by removing the SFC and TCR and rely on PFC and a real-time energy market.  The PFC is responsible for addressing fast power imbalances in timescales of tens of ms to few minutes (e.g., 100 ms to 5 minutes) while the real-time energy market is responsible for addressing longer imbalances in timescales of minutes to hours (e.g., 5 minutes to 1 hour).  TEC, on the other hand, is considered as optional.  
\end{abstract}

\vspace{2mm}


\section{Introduction} 
\label{sec:intro}



\subsection{Motivation}

Since the late 1920s, the structure of the frequency control has remained substantially unaltered.  It consists of real-time automatic primary frequency control (from speed governors), an automatic secondary frequency control (SFC) --- also known as automatic generation control (AGC) --- and an allocation of generation, based on economic considerations, known as tertiary control reserve (TCR).  In some cases, there can exist an automatic time error correction (TEC) that operates in a longer time scale than TCR.  These controls establish a hierarchal structure which is based on timescales separation.  
This structure is largely still in place today.  However, as a consequence of the move towards power systems with high shares of inverter-based resources (IBRs), the power system community has recognized the need to rediscuss needs, features and requirements of frequency control \cite{9286772, 8450880}.  In this context, we argue that the existing structure is not necessary in power systems with high penetration of IBRs, this leading to a simpler, yet equally secure, and potentially more economical operation of the grid.




\subsection{Background}

\begin{figure}[htb!]
  \begin{center}
    \resizebox{0.9\linewidth}{!}{\includegraphics{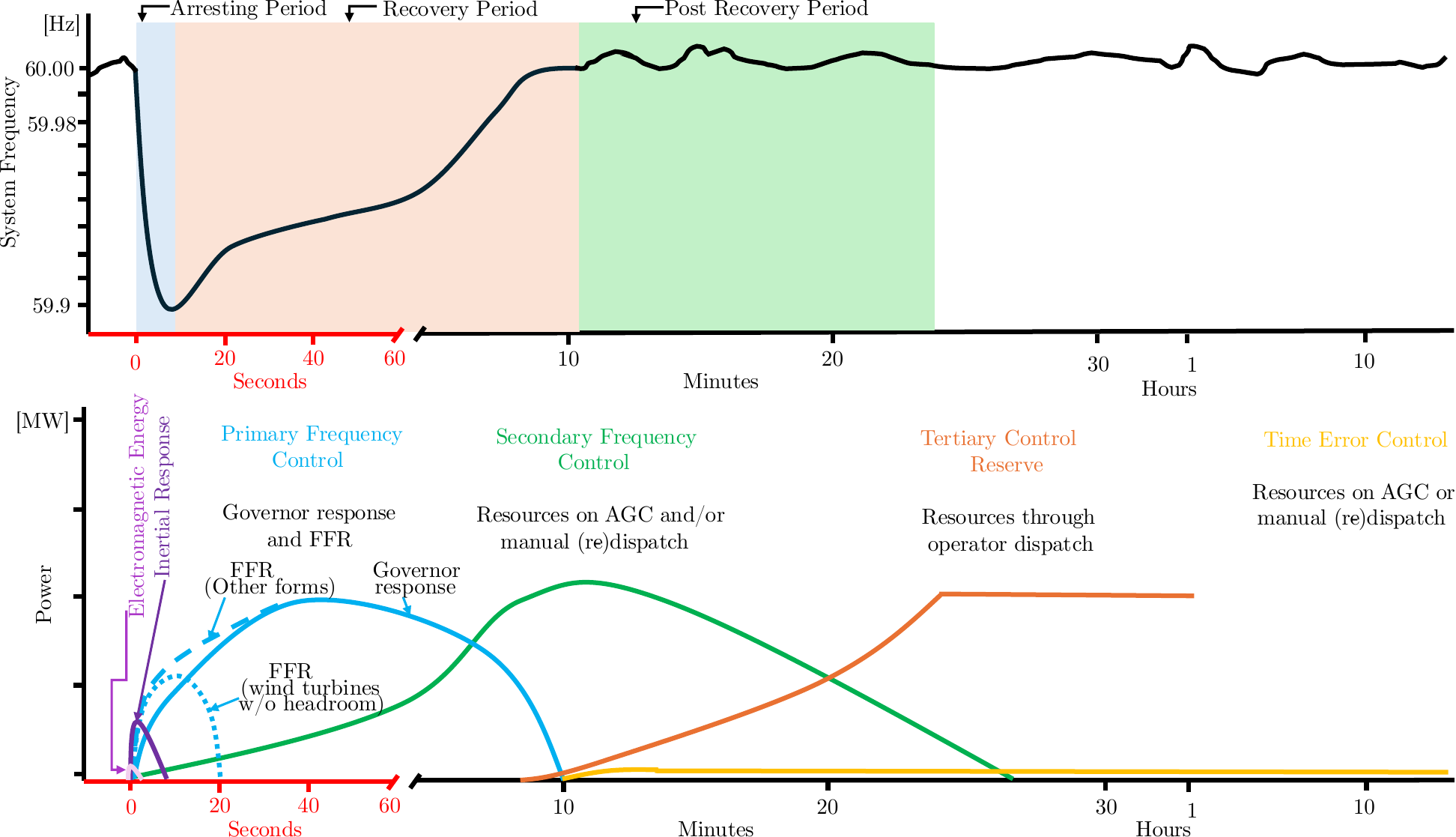}}
    \caption{Illustration of timescales of frequency control (adapted from \cite{9762253}).}
    \label{fig:timescales}
  \end{center}
  \vspace{-4mm}
\end{figure}

Figure \ref{fig:timescales} shows, assuming for illustration purposes an underfrequency (UF) event, that PFC (which includes fast frequency response (FFR)) is responsible to contain and recover frequency in timescales of seconds to several minutes (e.g., 5 minutes), then SFC is responsible to recover frequency to normal operating levels and remove the steady-state error following PFC provision in timescales of tens of seconds to tens of minutes (e.g., 5-15 minutes) and then TCR acts to replace SFC reserves in timescales of tens of minutes to one hour.  Finally, TEC makes sure that the accumulated time error does not exceed predefined limits such as $\pm$ 10 s or $\pm$20 s in timescales of hours to days.

\begin{figure}[htb!]
  \begin{center}
    \resizebox{0.9\linewidth}{!}{\includegraphics{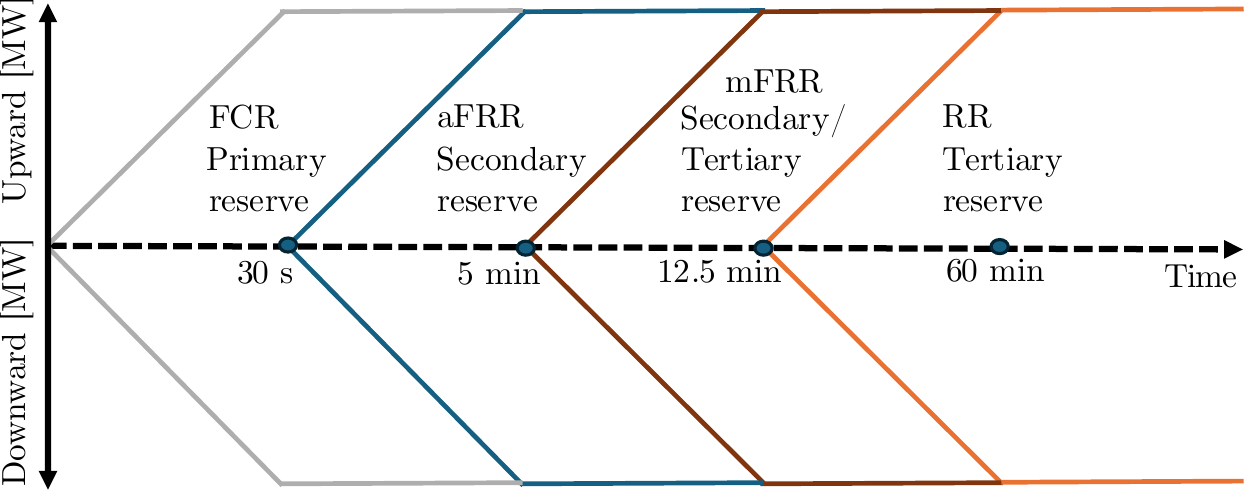}}
    \caption{Illustration of ENTSO-E standard products (adapted from \cite{nextreserves}).}
    \label{fig:pfcservices}
  \end{center}
  \vspace{-5mm}
\end{figure}

Based on the classification given in Fig.~\ref{fig:timescales}, transmission system operators (TSOs) have defined standard frequency reserve products to be procured in ancillary services markets.  For instance, Fig.~\ref{fig:pfcservices} depicts the different products defined by the European network of TSOs for electricity (ENTSO-E).  These reserves include frequency containment reserve (FCR) or PFC with a full activation time (FAT) of 30 s, frequency restoration reserve (FRR) or SFC including both automatic frequency restoration reserve (aFRR) with FAT of 5 minutes and manual frequency restoration reserve (mFRR) with FAT of 12.5 minutes.  The aFRR represents automatic SFC (AGC) while mFRR both manual SFC and part of manual TCR, and replacement reserve (RR) represents TCR.  The interested reader is referred to \cite{kercci2026frequency} and references therein for further detailed information on the hierarchical frequency control structure and the above reserves defined and used by ENTSO-E, respectively.  To avoid possible confusion, in the rest of the paper we will refer to PFC, SFC/AGC and TCR.

\subsection{Contributions}

This manuscript argues that SFC and TCR are not strictly necessary anymore in real-world converter-based power systems.  Based on this observation, the paper proposes to redesign the existing frequency control structure by removing the SFC and TCR and relying on PFC and a real-time energy market.  Specific novel contributions are as follows.
\begin{itemize}
    \item A comprehensive, evidence-based discussion, on why there is a need to rethink frequency control in real-world power systems.
    \item A new frequency control management approach that improves frequency quality, reduces power system complexity and potentially reduces ancillary services costs.
\end{itemize}
A case study based on a realistic model of the All-Island power system (AIPS) of Ireland supports the proposed new frequency control management approach.

\subsection{Organization}

The rest of the manuscript is organized as follows.  Section~\ref{sec:considerations} provides a detailed list of real-world evidence on why there is a need to rethink frequency control.  Section~\ref{sec:newtimescales} illustrates and discusses the proposed frequency control management approach and Section \ref{sec:market} provides a discussion on economic considerations related to it.  A case study based on a model of the AIPS is then presented in Section~\ref{sec:case}.  Finally, concluding remarks and future work directions are given in Section~\ref{sec:conclu}.


\section{Rethinking Frequency Control}
\label{sec:considerations}


This section discusses, based on practical considerations, pros and cons of existing control structure; the importance of mandatory stringent frequency control requirements; real-world evidence of PFC from different resources such as renewable energy sources (RES); distributed energy resources (DER) and high voltage direct current (HVDC) interconnectors (ICs); and their impact on different aspects of frequency such as frequency recovery, distribution, and maximum frequency deviations.\footnote{Some system operators rely on PFC from loads --- e.g., loads with controllable droop-type response in Hydro Qu{\'e}bec.  However, for space limitation, PFC from loads is not discussed in this section.}
The section also discusses, at a high-level, market design aspects.  


\subsection{Pros and Cons of Existing Frequency Control Structure}
\label{sec:proscons}

The advantages of the existing frequency control structure are certainly many and relevant.  Most importantly, it is relatively simple and is based on a century of operational experience.  It is thus well understood and has proven to work well under a large variety of operating conditions, including extreme events.  The hierarchical structure prevents control interactions and unwanted oscillations.  Moreover, it addresses frequency, tie-line load and TEC simultaneously and is substantially decoupled from energy markets and dynamics.  

On the other hand, existing frequency control structure has also several drawbacks.  The following is a list of the major issues that have been recognized by TSOs and practitioners:
\begin{itemize}
    \item Relatively slow for power systems that are increasingly experiencing faster dynamics.
    \item Difficult to adapt to novel technologies.  For instance, making AGC work using battery energy storage systems (BESS) with fast ramp rates and conventional units with relatively slow ramp rates is not straightforward.  
    \item Balances various objectives simultaneously (AGC) leading in some instances to opposite control signals \cite{opposite}.  
    \item SFC has communication delays.
    \item Designed based on large inertia and limited number of slow conventional synchronous generators. 
    \item Prone to cyber-attacks. 
    \item Generally neglects cost of power used for control. 
    \item No adaptive control and/or based on static parameters. 
    \item Includes single point failures due to centralized strategies. 
    \item Needs coordination and has low scalability and flexibility. 
    \item Relatively too many frequency control services/products. 
    \item Includes manual control. 
    \item Does not recognize developments in energy markets (e.g., real-time market with high time interval resolution). 
\end{itemize}

Furthermore, key parameters such as the control area's frequency bias characteristic (part of AGC), are difficult terms to quantify.  This is due to the fact that the frequency bias represents the combined droop characteristics of all the generators serving the load plus the frequency dependency of the load. However, as most AGC use integral controllers, this inability to accurately quantify the control area's frequency response characteristic has not caused particular issues \cite{493519}.

\subsection{Remarks on Mandatory PFC}
\label{sec:mandatory}

TSOs rely on minimum grid code requirements from transmission-connected resources to maintain system stability and security.  One critical requirement is the capability and ongoing enablement to provide mandatory PFC as part of the connection process.  For instance, the European Union legislation requires all transmission-connected resources, including RES and HVDC ICs, to be able to provide PFC with narrow deadband (e.g., $\pm$15 mHz) and PFC with wide deadband (e.g., $\pm$200 mHz), for both UF and OF.

In addition to mandatory requirements, and to make sure there is sufficient reserve available, system operators usually procure certain amount of balancing capacity through dedicated ancillary service products. Frequency control ancillary services/products work in tandem with mandatory PFC to ensure control frequency within and outside the standard frequency range (e.g., $\pm$200 mHz).  However, it is distinguished from mandatory PFC in the following key aspects:

\begin{itemize}
    \item 
    Ancillary services are provided by the reserved headroom, footroom, stored energy and defined frequency settings, as specified in a market ancillary service offer (e.g., based on day-ahead procurement).  Mandatory PFC, on the other hand, is a response based on a scheduled resource available capacity and energy at the time and with fixed frequency response settings.  
\item Ancillary services requirements may change continuously such as every 5 minutes in Australia (AUS), whereas mandatory PFC is an ongoing power system security requirement every time a resource receives a dispatch instruction to generate a volume greater than zero MW.

\end{itemize}

In addition, any frequency response provided within the standard range (e.g., $\pm$200 mHz) by an ancillary service provider providing mandatory PFC, is considered as contributing towards its delivery requirements for contingency ancillary service.  Another aspect to consider is the deviation from market dispatch due to mandatory PFC provision.  However, these changes tend to be small and short duration and will be minimised by having maximum participation in PFC. Thus, it will allow market participants to maintain close compliance with their energy market dispatch instructions.  For instance, when testing the implementation of narrow PFC deadbands in Tansmania (TAS) in 2018 the Australian Energy Market Operator (AEMO) stated: \textit{``As expected, narrowing the frequency deadband caused increased deviation in the generator’s output away from its dispatched load setpoint. The deviations were relatively small however and did not cross thresholds for energy dispatch non-conformance to be monitored''} \cite{taspfc}.

\begin{figure}[thb!]
  \begin{center}
    \resizebox{0.9\linewidth}{!}{\includegraphics{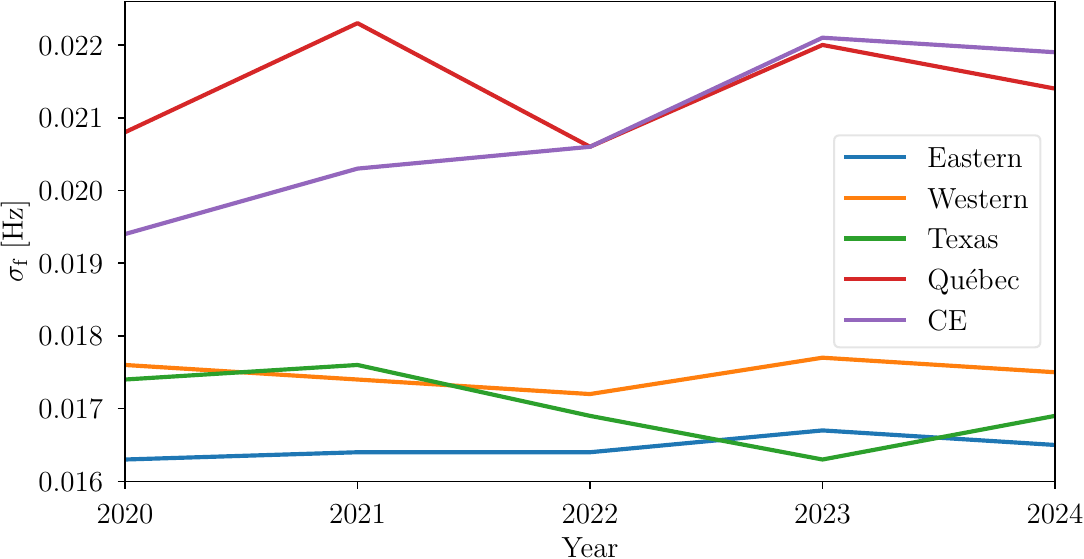}}
    \caption{$\sigma_{f}$ in recent years in CE and different North American systems \cite{nercercot}.}
    \label{fig:nerc}
  \end{center}
  \vspace{-4mm}
\end{figure}

In terms of benefits of mandatory PFC, it can enable the experimentation of new market structures or changes.  For example, because frequency control and related standards have been tightly enforced by the North American Electric Reliability Corporation (NERC), see Table \ref{tab:nercdbdroop} \cite{nercdb}, the actual performance of the interconnected power systems has not been affected by experimentation with new market structures \cite{1259344}.  That is, the overall frequency performance has been relatively good because of the very stringent frequency standards.  For instance, the Electric Reliability Council of Texas (ERCOT), the Texas system operator states that: \textit{``ERCOT saw big improvement in its primary frequency capability following implementation of BAL-TRE-001 Regional NERC Standard.''} \cite{ercotpfc}, which required, among others, that mechanical and electronic and digital governor deadbands of all generators to be no greater than $\pm$36 mHz and $\pm$17 mHz, respectively.   This is not the case, for example, in the Continental European (CE) power system where TSOs implement less stringent PFC requirements compared to those in North America.  This is, in turn, one of the main reasons the quality of frequency control in the CE under normal operating conditions has been steadily degrading (last 20+ years).  Figure~\ref{fig:nerc} shows the evolution of the standard deviation of the frequency, $\sigma_{f}$, for CE and different North American systems.  For comparison, the Texas power grid shows steady and lower $\sigma_{f}$ than CE despite being around 5 times smaller (see Table \ref{tab:different}) and with much higher IBR penetration levels.  As a matter of fact, in 2023, CE exceeded for the first time the annual target of less than 15,000 minutes outside $\pm$50 mHz range (15,389 minutes).

\begin{table}[h!]
  \centering
  \caption{PFC requirements for CE and different North American systems.} 
  \label{tab:nercdbdroop}
  \resizebox{1.0\linewidth}{!}{
  \begin{tabular}{lccccccc}
    \hline
     Power System & Max droop & Max db  & Narrow db  & Narrow db \\
     & (\%) & ($\pm$mHz) & at all times? & for all resources? \\
     \hline
    Eastern & 4--5 & 36 & Yes & Yes \\
    Western  & 4--5 & 36 & Yes & Yes \\
    Texas & 4--5 & 17--34 & Yes & Yes \\
    Québec  & 5 & 0 & Yes & Yes
  \\
  CE  & 12 (default 4-5) & 500 & No & No \\
   &  & 10 (PFC) \\
      &  & 200 (PFC) & &  \\

    \hline
  \end{tabular}}
\end{table}

Hence, frequency quality is expected to deteriorate if TSOs integrate more and more uncertain, variable and inertialess RES without requiring these resources and other relevant resources such as BESS to operate at all times with narrow PFC deadband.  Other relevant measures that TSOs could consider include:
\begin{itemize}
\item increase volumes of procured reserves; 
\item change frequency response settings (e.g., lower droop \%).
\end{itemize}

A scenario where the provision of mandatory PFC is particularly critical is during severe system conditions such as system splits.  Indeed, dealing with system splits is the main concern of large interconnected power systems such as the CE rather than dealing with single contingencies (i.e., N-1).  One could argue that part of the reason of the Chilean and Iberian blackouts in 2025 was the lack of mandatory frequency and voltage support for the vast majority of resources.  In other words, despite the CE dimensioning PFC needs based on a N-2 criterion (3 GW) and not N-1 \cite{entsoen2}, this volume is proportionally shared within the CE synchronous area where the Spanish and Portuguese TSOs procure only a small share of 3 GW (e.g., around 300-400 MW).  But if mandatory PFC was in place then in the system split scenario the TSOs would have had much more reserves to deal with the situation.  

For instance, following the system separation event in AUS on 25/08/2018 where Queensland and South AUS regions were separated from the remainder of the National Electricity Market (NEM) AEMO noted \cite{aemosplit}: \textit{``This event indicates that the resulting decrease in primary frequency control has significantly reduced the ability of the power system to arrest the impact of non-credible contingency events in time to avoid the risk
of cascading failures''}, and made the primary recommendation of: \textit{``increasing the provision of primary frequency control from capable generation to arrest
the decline in system resilience to larger contingency events and maintain frequency closer to 50 Hz''}.  These excerpts suggest that stringent
requirements for frequency control are needed to enforce ``grid discipline'' and, in turn, maintain good or excellent frequency performance and stability (same thing can be said for voltage control).  In other words, mandating frequency control for a diverse set of resources substantially increases the number of facilities that are tightly controlling frequency which, in turn, makes the grid more resilient.  

Of course, the economics need to be kept in mind as strict mandatory frequency control requirements, such as those implemented in North America, may lead to an over-consumption that can result in higher costs to market participants which, in turn, can
result in higher costs to consumers.  For instance, generators may not be able to recover costs including wear and tear caused by continual governor activity due to PFC with narrow deadband.
 This same argument has been raised by industry in AUS after the introduction of mandatory PFC in 2020.  To address such concerns from industry, the AUS regulatory authority (RA) and TSO decided to implement frequency performance payments that include financial incentives and penalties to encourage facilities such as generators, large loads and BESS to operate their plant in a way that helps to control
frequency \cite{nempfc}.  In this context, we think that it is prudent that power system needs should come first. Then, based on those identified needs, the TSOs, along with the RAs, can focus on how to best design energy / ancillary frameworks to meet those needs from an economic perspective.


\subsection{PFC from Large-Scale RES}
\label{sec:res}

\begin{figure}[thb!]
  \begin{center}
    \resizebox{0.85\linewidth}{!}{\includegraphics{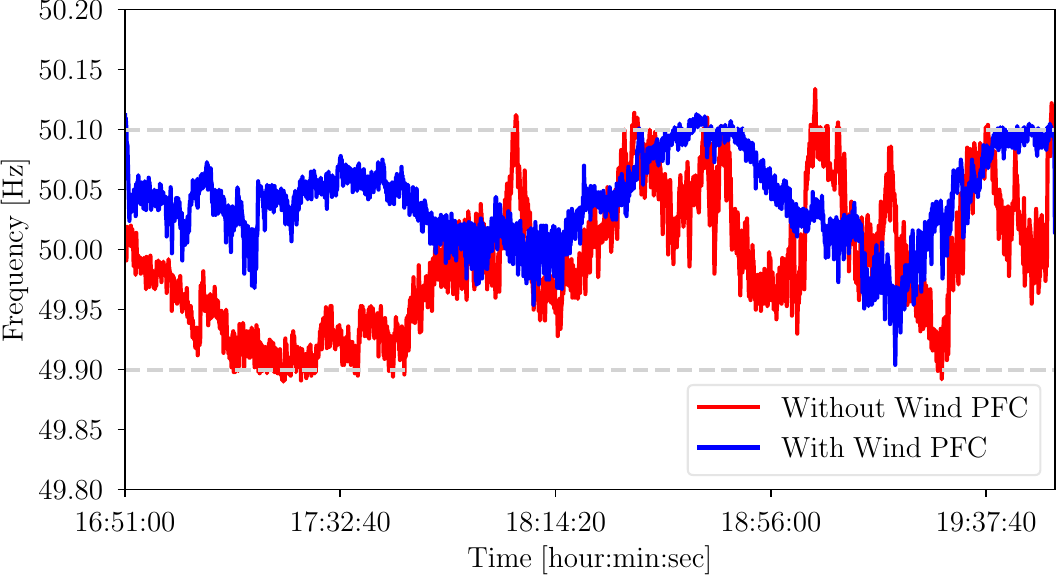}}
    \caption{Frequency traces of the AIPS with \& without wind PFC.}
    \label{fig:apc}
  \end{center}
  \vspace{-4mm}
\end{figure}

In this section, we discuss a real-world example of large-scale wind power plants providing PFC with $\pm$15 mHz deadband.  More specifically, Fig.~\ref{fig:apc} shows frequency traces from 27/01/2024, namely 6 consecutive hours, 3 h when the PFC deadband was set $\pm$200 mHz and 3 h when the PFC deadband was set $\pm$15 mHz.  During this period, the AIPS experienced high wind generation of around 3.5 GW.

Figure~\ref{fig:apc} shows that when wind generation provides PFC with $\pm$15 mHz deadband, overall frequency performance improves.  This is particularly visible at the beginning of the figure where frequency becomes less volatile.  In turn, this means that when using narrow deadbands on RES control room operators spend less time issuing manual dispatch instructions (e.g., to conventional generating units) to keep frequency within the $\pm$200 mHz range.  Observe that the AIPS does not utilize an AGC.  Also, note that if wind is not pre-curtailed it will only provide downward regulation.  Of course, enabling narrow deadbands means wind is being curtailed more to provide balancing services and thus induces a cost to wind farm owners and, ultimately, to end consumers.  But if this balancing service is incentived TSOs could increase the flexibility of their systems and better maximize the integration of RES while achieving economic benefits and having more reserves available.  However, as demonstrated in \cite{10849615}, this type of PFC from wind introduces asymmetry in the frequency distribution.  This issue is discussed in Section~\ref{sec:pfc dist}.  

\subsection{PFC from DER}
\label{sec:der}

The reliable PFC provision from DER, including through the virtual power plant concept, is critical for secure operation of current and future power systems \cite{ZHONG2020106609}.  Real-world testing such as those in \cite{rooftop} demonstrate that roof-top photovoltaic (PV) systems are able to provide PFC services.  It is important that these capabilities are enforced into national grid codes to manage relevant scenarios such as blue-skies where, for example, most of the demand is being met by distributed roof-top PV.  As a matter of fact, in Europe, new revised network codes on requirements for generators and demand connection code will require DER such as, for example, distributed PV and electric vehicles and heat pumps to provide mandatory OF and UF PFC services.  Bidirectional PFC from DER is also enforced in AUS, where they are capable of doing so.

\subsection{PFC from HVDC Interconnectors}
\label{sec:hvdc}

Modern HVDC links have the capability to operate with narrow deadbands and thus greatly support frequency regulation.  In fact, it is shown in the literature that making HVDC ICs frequency supportive is also cost effective \cite{hvdcnordic}.  However, their PFC contribution is often overlooked or not known by power system engineers and researchers.  In this section, we provide a real-world example from TAS,  where the HVDC link between TAS and AUS is a major source of frequency regulation.  It should be noted that while older HVDC links may not be suitable for continuous power changes due to wear and tear, modern HVDC ICs have this issue less and so can provide significant more frequency support.  In addition, it is also worth mentioning that future grid-forming (GFM) HVDC ICs will inherently have superior frequency control capabilities.

\begin{figure}[thb!]
  \begin{center}
    \resizebox{0.85\linewidth}{!}{\includegraphics{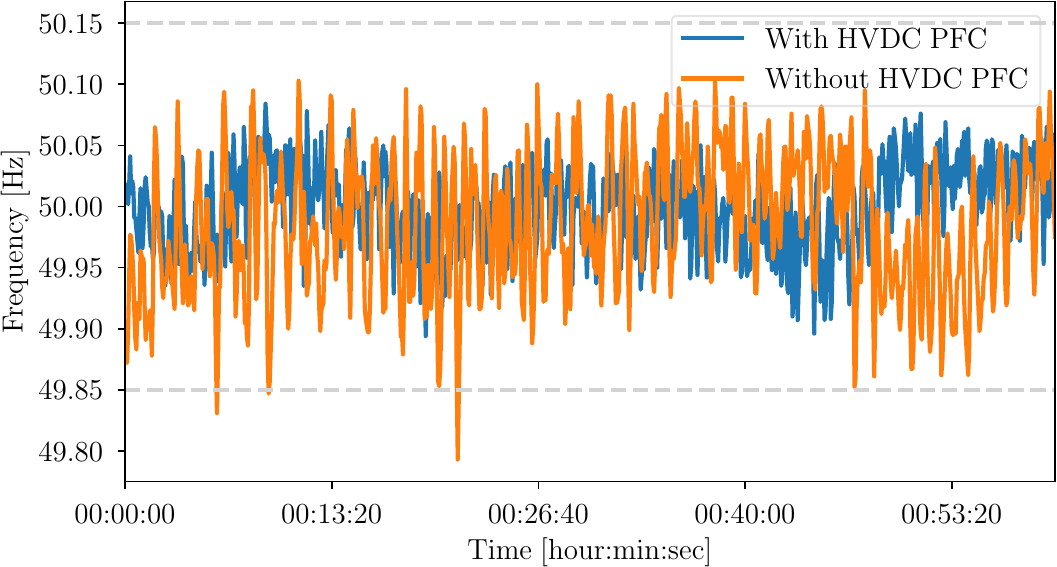}}
    \caption{Frequency traces for one hour in the TAS power system with (08:30 a.m to 09:30 a.m) and without (01:30 a.m to 02:30 a.m) HVDC PFC.}
    \label{fig:tas}
  \end{center}
  \vspace{-4mm}
\end{figure}

Figure~\ref{fig:tas} depicts two 1 h 2026 frequency traces of TAS from 23/02/2026 namely one where the HVDC link provides PFC and one where it does not.  It is clear that when the HVDC link does not provide PFC, the frequency is much more volatile and spends quite some time outside the standard frequency range namely $\pm$150 mHz.  Note that in both cases AGC is enabled in the system.  This means that the effectiveness of AGC is negligible compared to the contribution of PFC and, in this case, PFC of only one service provider.  

\subsection{Impact of PFC on Frequency Recovery}
\label{sec:pfc recovery}

An important aspect of frequency control is how quickly it can recover frequency within normal operating bands such as $\pm$150 mHz for AUS following credible contingencies.  We illustrate the impact of PFC and, in particular, mandatory PFC on frequency recovery using operational data from the AUS.  Specifically, we select similar contingencies in 2018 (without mandatory PFC) and 2025 (with mandatory PFC) namely trip of the Loy Yang B No. 1 generating unit from 530 MW at 15:28 on 18/01/2018 and trip of Loy Yang B power station unit 2 from 579 MW at 15:41 on 12/01/2025, respectively.

\begin{figure}[thb!]
  \begin{center}
    \resizebox{0.85\linewidth}{!}{\includegraphics{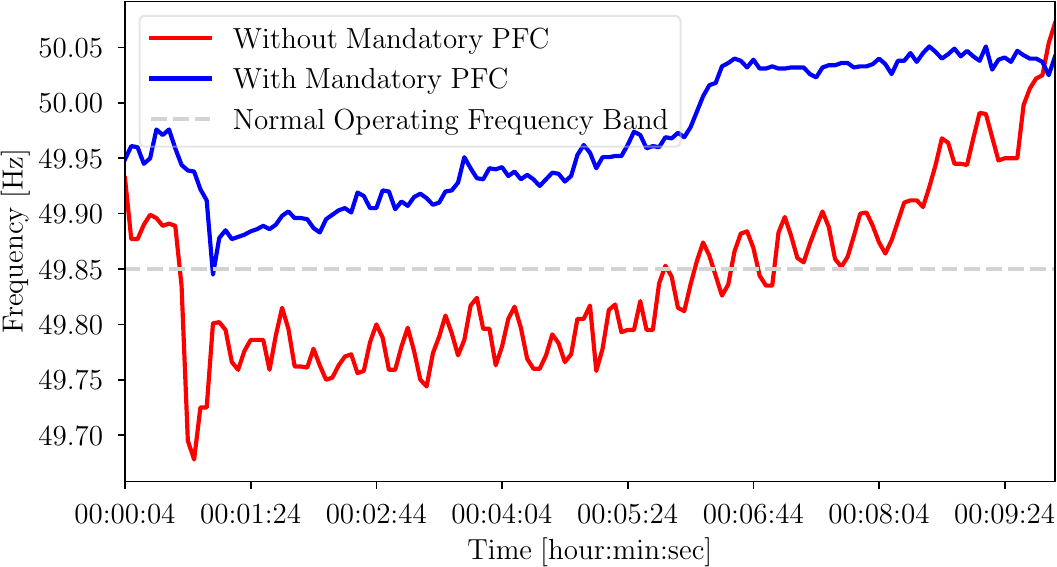}}
    \caption{Frequency traces for the AUS system with \& without mandatory PFC.}
    \label{fig:recovery}
  \end{center}
  \vspace{-4mm}
\end{figure}

Apart from some differences in the operating conditions, Fig.~\ref{fig:recovery} clearly shows that the mandatory PFC rule has contributed to a rapid recovery of frequency within the normal band and toward 50 Hz compared to the slow recovery in the case of without mandatory PFC.  In fact, AEMO reports that: ``\textit{System frequency is now recovering to within the normal operating frequency band within 10 to 20 seconds for large credible contingency
events due to abundant PFR, which is faster than AGC’s typical control cycle, meaning AGC plays only a minor role in
these shorter frequency deviations.''} \cite{aemorecovery}.  Such improvement in the dynamic performance is expected to continue if we consider the significant increase in the number of resources capable of providing fast and accurate PFC.  Therefore, it is fair to say that based on real-world evidence having an AGC to deal with contingency events for frequency recovery purposes does not seem relevant anymore.  In fact, it has been shown in the industry that the fast response from IBRs also enables a reduction in required MW reserves  \cite{ercot}.

\subsection{Impact of PFC on Maximum Frequency Deviations}
\label{sec:pfc nadir/zenith}

With regard to maximum frequency deviations,  Fig.~\ref{fig:recovery} shows that for similar contingency sizes mandatory PFC in AUS has led to an improved, in this case, frequency nadir.  This has to be expected given the aggregate frequency response available, in particular, from fast-acting resources such as BESS.  For instance, AEMO anticipates approximately 20 GW of BESS by the early 2030s.  If this is the case, it means that 20 GW of
BESS with 1.7\% droop (requirement) and $\pm$15 mHz PFC deadband applied will provide 3.176 GW of response by the time frequency reaches the normal operating frequency band threshold of 50$\pm$150 mHz, as given by \cite{aemorecovery}:
\begin{equation}
  \label{eq:aggregate}
  \rm Freq. \; Resp. = \frac{100}{Droop \; \%}\cdot \frac{\Delta f}{50}\cdot BESS_{total} = 3.176 \; GW \, ,
\end{equation}

\begin{figure}[thb!]
  \begin{center}
    \resizebox{0.8\linewidth}{!}{\includegraphics{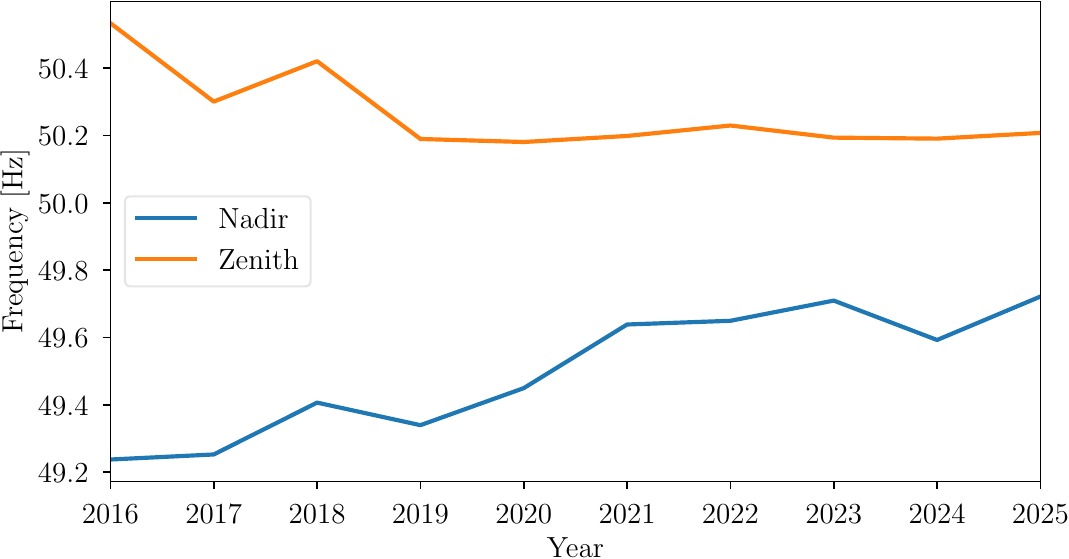}}
    \caption{Evolution of maximum instantaneous frequency deviations in the AIPS.}
    \label{fig:nadir_zenith}
  \end{center}
  \vspace{-4mm}
\end{figure}

A similar calculation shows the same BESS fleet would provide 235 MW for every 0.01 Hz beyond their $\pm$ 15 mHz PFC deadband.  Taking into account that sudden changes in the supply-demand balance in AUS rarely exceed 700-800 MW at present (i.e., represent largest single infeed loss), this suggests that the largest credible generation contingencies will move system frequency by approximately $\pm$ 30 mHz, which is slightly higher than PFC deadband.  Therefore, this simple theoretical example also shows that managing frequency with PFC should be feasible in current and future converter-based power systems.  This statement is further supported by operational data from the AIPS (see Fig.~\ref{fig:nadir_zenith}) where the maximum frequency deviations have improved in recent years despite the increase in IBR penetration.  The main reason for this improvement is increased PFC provision from BESS units.  These real-world evidences suggest that once PFC reaches some critical mass of MW/0.1 Hz, relative to typical MW contingency sizes, SFC starts to become unnecessary as there will not be enough of a sustained frequency error for AGC to play a useful role in correcting the error.  And with IBRs, especially BESS, on low frequency droop settings, this is what is expected to happen in high IBR systems.

\subsection{Impact of PFC on Frequency Distribution}
\label{sec:pfc dist}

\begin{figure}[htb]
  \subfigure[GB, AIPS, AUS, TAS, CE, Nordic]{\resizebox{0.49\linewidth}{!}{\includegraphics{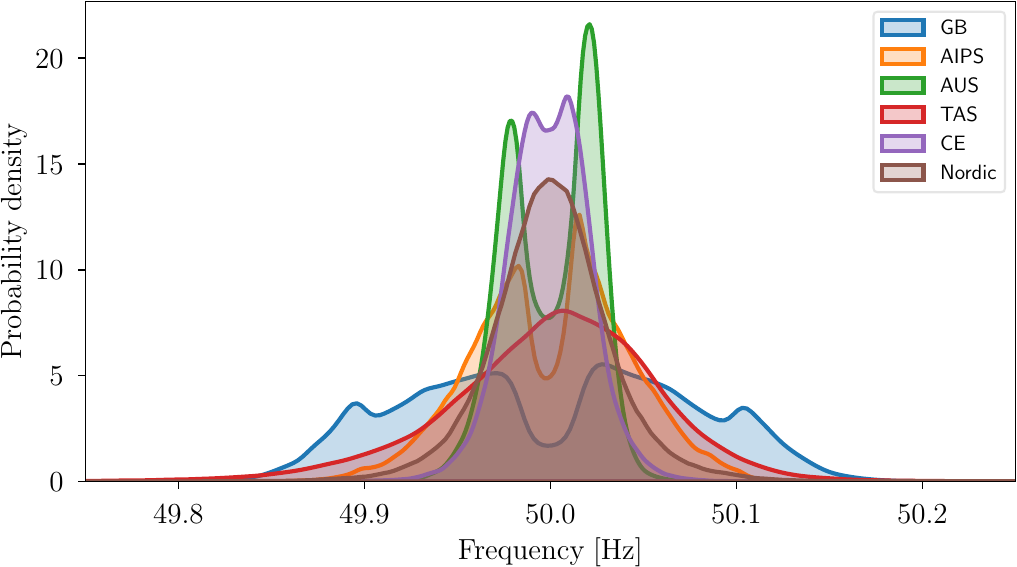}}}
  \subfigure[Texas]{\resizebox{0.49\linewidth}{!}{\includegraphics{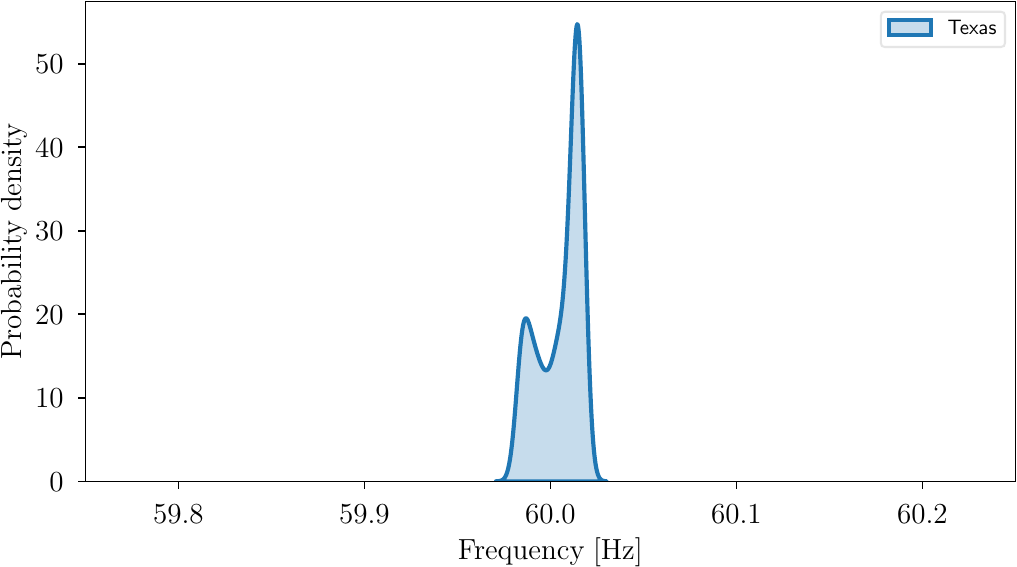}}} \\
  \caption[Frequency distributions for different power systems.]{Frequency distributions for different power systems.
  }
  \label{fig:dist}
\end{figure}

Here, we discuss the impact of PFC, including mandatory PFC, on the frequency distribution and, in particular, on asymmetry of frequency distribution which is a concern for some TSOs \cite{10849615}.  With this aim, Fig.~\ref{fig:dist} shows the frequency distribution for different power systems namely Great Britain (GB), AIPS, AUS, TAS, CE, Nordic and Texas.  For GB, AIPS, AUS, TAS, CE, and Nordic we use September 2025 frequency time series data with 0.1s, 1s and 4s resolutions while for ERCOT we utilize 2h data with 10s resolution from 22/01/2026 (still sufficient to illustrate impact of PFC).

\begin{table*}[t!]
  \centering
  \caption[Relevant characteristics of different selected power systems]{Relevant characteristics of different selected power systems.}
  \label{tab:different}
  \resizebox{0.8\linewidth}{!}{
  \begin{tabular}{cccccccc}
    \hline
    Item & GB & AIPS & AUS & TAS & CE & Nordic & Texas\\
    \hline
    Peak demand [GW] & 44 & 7.5 & 34 & 2  & $>$ 400 & 60 & 85\\
    PFC provision & Market/Mandatory & Market/Mandatory & Mandatory & Mandatory & Market/Mandatory & Market/Mandatory & Market/Mandatory\\
    deadband [mHz] & $\leq \pm 15$ & $\leq \pm 15$ & $\leq \pm 15$ & $\leq \pm 15$ & $\leq \pm 10$ & $\leq \pm 10$ & $\leq \pm 17$ \\
    Droop [\%] & 3-5 & 3-5 & $\leq$ 5 & $\leq$ 5 & $\leq$ 5 & $\leq$ 5 & $\leq$ 5 \\
    FFR & Yes (FAT = 1 s) & Yes (FAT = 0.15-2 s) & Yes (FAT = 0.5-1 s) & Yes (FAT = 0.5-1 s) & Limited & Yes (FAT = 0.7-1.3 s) & Yes (FAT $\approx$ 0.25 s)\\
    AGC & No & No & Yes (FAT = 5 mins) & Yes (FAT = 5 mins) & Yes (FAT = 2-15 mins) & Yes (FAT = 5 mins) & Yes (FAT = 5 mins) \\
    Dispatch model & Self & Central & Central & Central & Mainly self & Self & Central\\
    \hline
  \end{tabular}}
\end{table*}

It is interesting to see that while no power system shows exactly the same frequency distribution, 6 out of the 7 selected systems (i.e., except the Nordic) show a bi-modal distribution.  A possible explanation on why the Nordic system shows a normal distribution could be related to the fact that being a hydro-dominated system it means that water column physics smear the effect out.
It is worth mentioning that while all TSOs require mandatory PFC provision from all resources, not all of them operate all resources with narrow deadband (e.g., $\pm$10-15 mHz).  For instance, TSOs in Europe generally do not operate RES with narrow deadbands but rather with wide ones, that is, with PFC deadband of $\pm$200 mHz.  When frequency is outside $\pm$200 mHz resources contracted to provide the market service with narrow deadband can remain in the same mode of operation, but all other resources need to provide PFC using $\pm$200 mHz deadband (mandatory).  As discussed above, such requirements differ from those in North America and AUS where all resources, including wind, solar and BESS, are mandated to operate at all times with a narrow PFC deadband.


As discussed in \cite{10849615}, the main source of this asymmetry is RES such as wind and solar power providing PFC with narrow deadband such as $\pm$10 -- 17 mHz.  Other relevant sources include control limits and network losses.  Note that frequency spends less time below the deadband due to the fact that most wind and solar resources are being operated at maximum output (through the maximum power point tracking controller) leading to an inability to increase output when frequency falls below the deadband.  Nevertheless, it is worth recognizing that while narrow PFC deadband on RES introduces asymmetry, it also allows keeping frequency within a narrow band.  For instance, despite the significant asymmetry, the Texas power grid shows that frequency stays very close to 60 Hz.  This is due to the fact that the \textit{``Standard BAL-001-TRE-112 requires all resources in the Texas Interconnection to provide proportional, non-step
primary frequency response with a ±17 mHz. dead-band. As a result, if any time frequency exceeds 60.017 Hz,
resources automatically curtail themselves. That has resulted in far less operation in frequencies above the deadband since all resources, including wind and solar, are backing down''} \cite{nercercot}.  The fact that frequency stays around the PFC deadband means the effect of AGC is neglible.  This statement is in agreement with relevant literature such as \cite{8626538} where it is shown that the PFC deadband width and aggregate system droop are
the major parameters determining $\sigma_{f}$. 

\begin{figure}[thb!]
  \begin{center}
    \resizebox{0.7\linewidth}{!}{\includegraphics{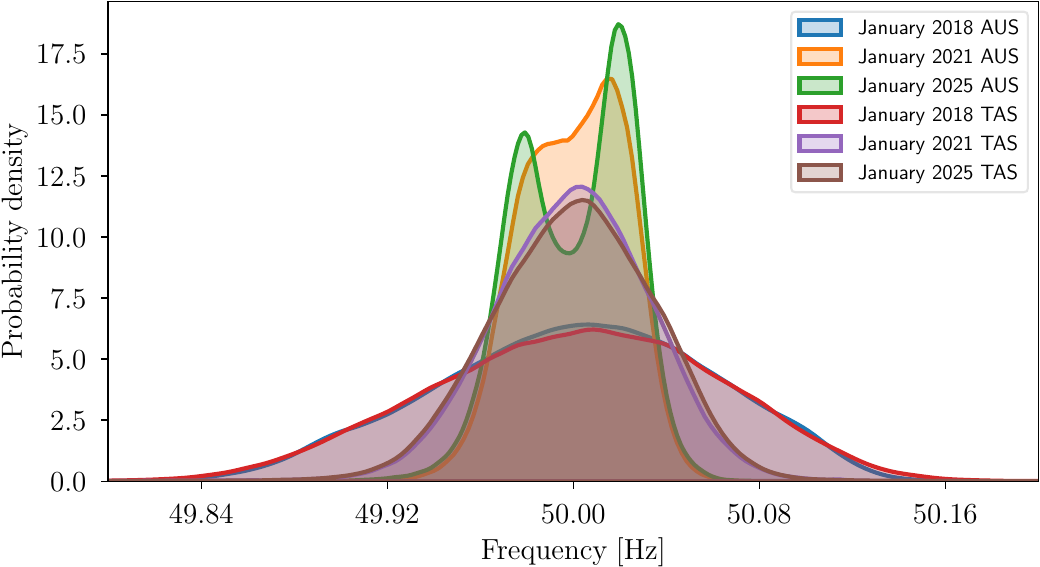}}
    \caption{Frequency distributions for 2018, 2021, and 2025 for AUS and TAS.}
    \label{fig:aus2}
  \end{center}
  \vspace{-4mm}
\end{figure}

Mandatory PFC with narrow deadband is also the main reason why frequency is staying within a narrow band in AUS and TAS as shown in Fig.~\ref{fig:aus2}.  However, in contrast to AUS, asymmetry does not seem to have increased in TAS despite the fact that the mandatory PFC deadband rule was introduced in 2020 in both AUS and TAS \cite{11299077}.  This could be explained by the excellent and, apparently, symmetric frequency response provided by the HVDC link in the TAS system as seen in Fig.~\ref{fig:tas}.  Note that both TAS and AUS utilize an AGC.  This means that the AGC effectiveness when it comes to reducing asymmetry of frequency distribution is negligible compared to, for example, PFC from HVDC.  Therefore, a solution to asymmetry could be, where possible, operating HVDC links with narrow deadbands.  GFM BESS might also be a good candidate to reduce the asymmetry as shown in \cite{kerci2025frequency}.  


\subsection{Frequency Control Performance}
\label{sec:smallvslarge}

\begin{figure}[b!]
  \begin{center}
    \resizebox{0.85\linewidth}{!}{\includegraphics{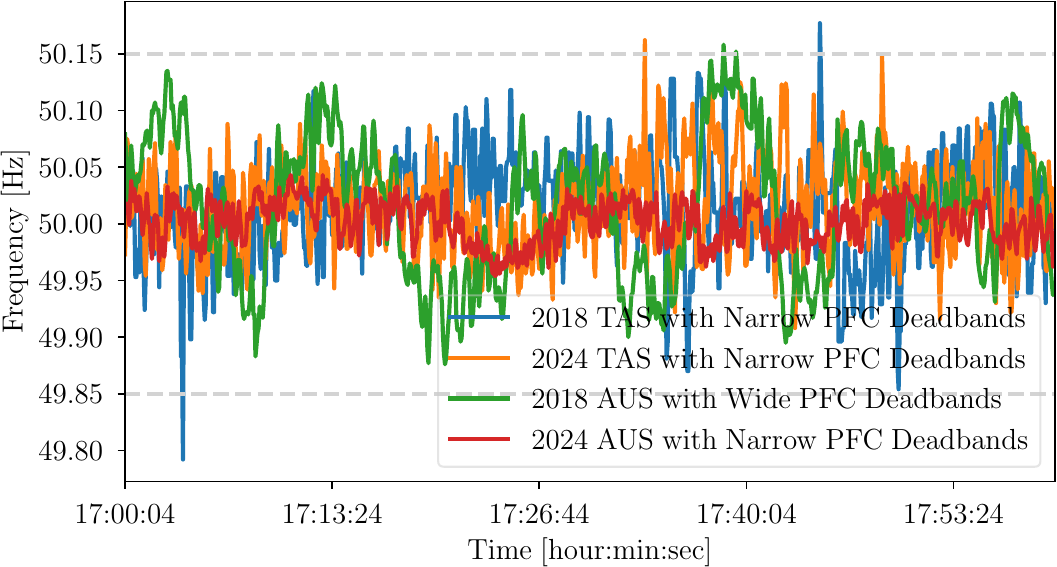}}
    \caption{Frequency traces for TAS and AUS in 2018 and 2024.}
    \label{fig:smallvslarge}
  \end{center}
  \vspace{-4mm}
\end{figure}

To illustrate the importance of operating with narrow PFC deadbands we compare the frequency control performance of a small island system such as TAS, with a large one such as mainland AUS.  With this aim, we select 1 h from the same day namely 17 May from 17:00--18:00 in 2018 and 2024 in TAS where it was operating with narrow PFC deadbands and in 2018 and 2024 in AUS where it was operating with wide (i.e., $\pm$150 mHz) and narrow (i.e., $\pm$15 mHz) PFC deadbands, respectively.  Figure~\ref{fig:smallvslarge} shows the relevant frequency traces.  

The comparison suggests that size of power systems alone may not be an important factor but rather width of the frequency deadbands, the response capabilities of the generation/demand resources as well as the level of MW noise/sizzle the system experiences relative to its size are equally important as well.   For instance, TAS generally shows similar or better frequency performance than mainland AUS when the latter was applying wide PFC deadbands in 2018 despite being more than 15 times smaller than AUS.  The figure also shows that when AUS was operating with narrow deadbands in 2024 following the mandatory PFC rule in 2020 then, as expected, frequency variations are smaller in bigger systems (2024 AUS) than small systems (2024 TAS).  Note that in all cases AGC was enabled.  These real-world observations challenge the general understanding that small island power systems face greater frequency control challenges than large systems (i.e., due to lower inertia) and instead place the importance onto how actually TSOs and the respective RAs design and implement frequency control arrangements, including mandatory ones.

\subsection{Need for Perfect Tracking Frequency Controller}
\label{sec:pi}

An obvious question for systems operating without a perfect tracking integral controller (i.e., AGC) is whether they can be operated securely or how can frequency be managed without such a controller.  Real-world data presented in this paper show that it is possible to operate power systems securely, including low-inertia ones, without a perfect tracking controller from a frequency perspective.  Note that AGC may be used for other purposes such as, for example, for market dispatch purposes where 5-minute dispatch results are ramped into the AGC to prevent any large step
change in MW output (i.e., AUS).  However, this aspect is not strictly related to frequency control and, thus, is outside the scope of the paper.
Moreover, one may argue that even today AGC is not really doing perfect tracking of system frequency deviations as system is never in steady state due to continuous load and generation fluctuations and slow AGC timescales (e.g., 5 minutes FAT).






\subsection{Remarks on TEC, TCR and Tie-Line Control}
\label{sec:tfc}

TEC, including automatic TEC through AGC, is still widely performed by TSOs worldwide.  However, in recent years some jurisdictions such as AUS have removed the regulatory obligation to correct time error to within certain bands.
NERC has followed a similar path with the retirement of NERC Standard BAL004-0 meaning that \textit{``TEC is no longer initiated when the time error crosses predefined thresholds. However, TEC
may still be initiated if needed for reliability concerns or other unforeseen reasons''} \cite{caiso}.  Therefore, it is fair to say that TEC is becoming less and less relevant for time keeping and in the future it may be completely removed as there are fewer devices relying on it.  But even if it will be required (e.g., as a useful diagnostic for MW balance), its long timescales, that is, hours to days, means that it will be a relatively easy task for TSOs (e.g., intentionally dispatching more or less generation).

Regarding TCR, AUS have run a 5-minute MW dispatch cycle since 1998, which serves the role of TCR.  This has allowed AUS to avoid doing MW tie-line control between areas in the AUS grid, as 5 minutes was considered short enough that tie-line MW errors could be corrected via re-dispatch at the next run.  This has, in turn, allowed AUS to run the system as a single flat frequency control area under normal conditions, which greatly simplifies ancillary services market operation.  

With regard to tie-line control, while it is embedded into a frequency control process namely AGC, strictly speaking, it is a problem of load control rather than frequency control.  This was stated by researches since its beginning around a century ago: \textit{``Regulation of tie lines interconnecting two electric power systems is a problem of load control rather than one involving principally system frequency; for best results, the control equipment must be automatic.}'' \cite{6429761}.  Being a load control problem, it means it could be managed through real-time security-constrained economic dispatch markets.  Excellent frequency control would help tie-line control as well.

\section{Proposed Frequency Control Structure}
\label{sec:newtimescales}







\begin{figure}[thb!]
  \begin{center}
    \resizebox{0.9\linewidth}{!}{\includegraphics{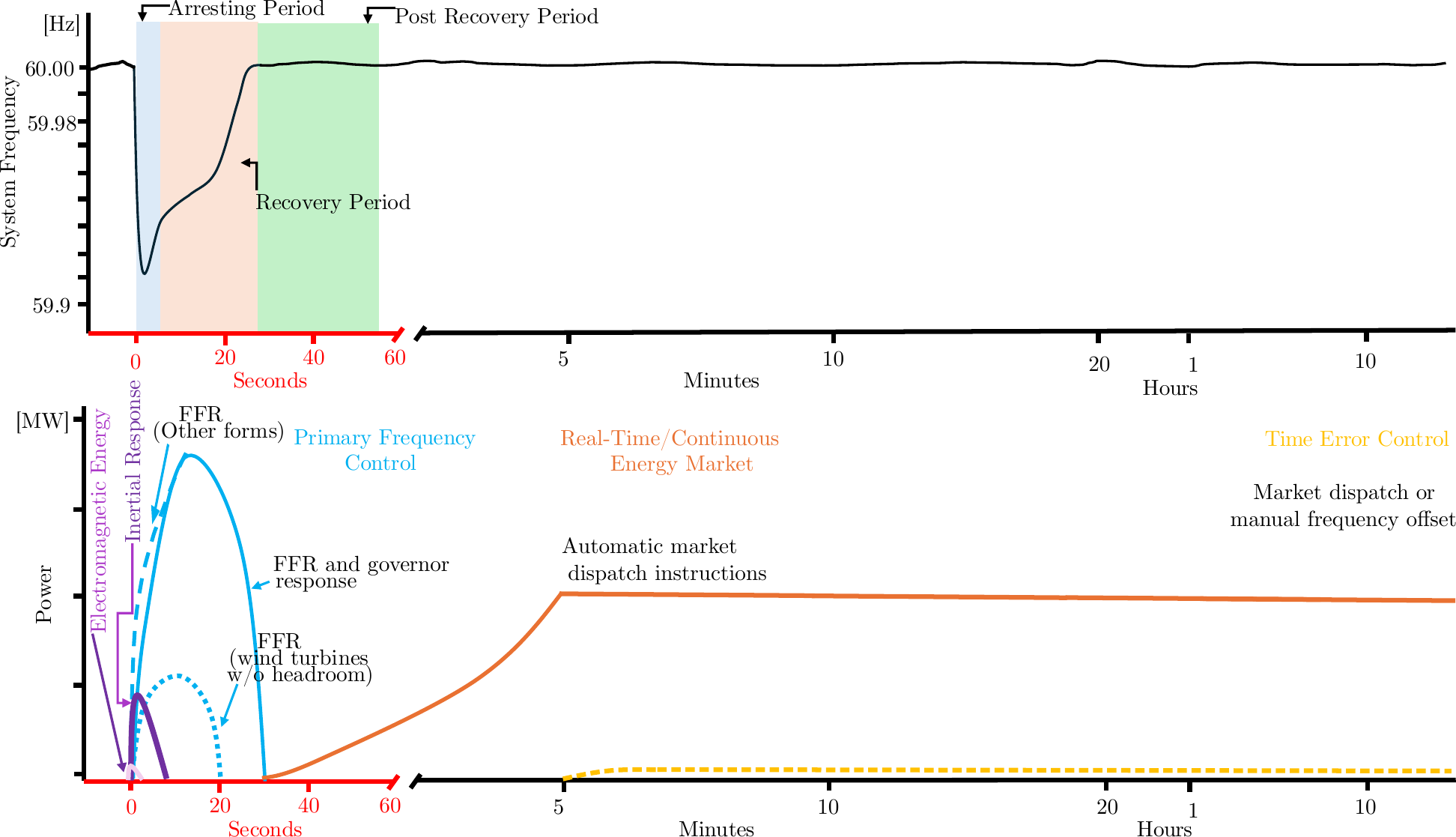}}
    \caption{Illustration of new timescales of frequency control.}
    \label{fig:newtimescales}
  \end{center}
  \vspace{-4mm}
\end{figure}

\begin{figure}[thb!]
  \begin{center}
    \resizebox{0.85\linewidth}{!}{\includegraphics{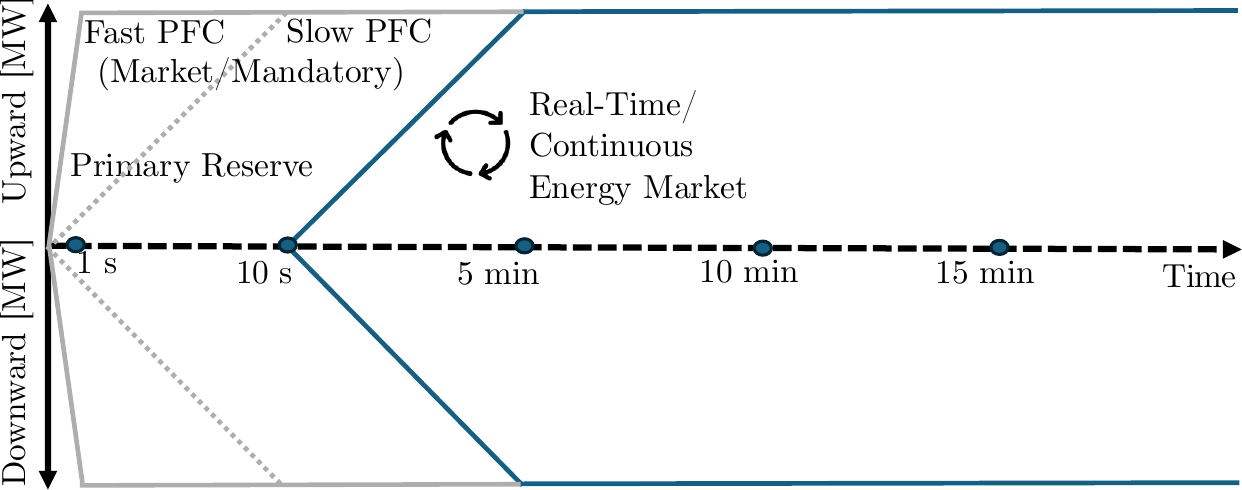}}
    \caption{Illustration of proposed new frequency control management concept.}
    \label{fig:newpfcservices}
  \end{center}
  \vspace{-4mm}
\end{figure}

Figures~\ref{fig:newtimescales} and \ref{fig:newpfcservices} illustrate the envisioned frequency control timescales and management in power systems.  We can now see that, as opposed to Fig.~\ref{fig:timescales}, where frequency recovers to nominal within 10 minutes, it does so now within 30 s following the contingency due to fast and abundant PFC.  The latter is also the reason why frequency nadir has improved.  Additionally, it displays a tighter regulation around the nominal during frequency post recovery period and in steady-state due to fast and abundant PFC and, to a lesser extent, due to efficient real-time energy markets able to keep good power balance close to physical delivery.  

In contrast to Fig.~\ref{fig:pfcservices} that includes PFC, SFC and TCR, Fig.~\ref{fig:newpfcservices} shows that power systems could be operated securely by relying on strict mandatory PFC requirements similar to those in AUS and North America, ancillary services markets that properly address any gap between mandatory PFC and acceptable frequency quality levels and, last but not least, efficient real-time energy markets.  Note that depending on the resource mix and market structures TSOs may implement different PFC reserve dimensioning approaches (e.g., covering largest single infeed/outfeed losses plus additional PFC reserves to tackle imbalances resulting from forecast errors).  Moreover, it is worth mentioning that the PFC and market timescales are for illustrative purposes and are not exact as they may not work for all power systems. This is the case even today for existing standard reserve products in Fig.~\ref{fig:pfcservices} where, for example, a PFC with 30 s FAT is (still) working for CE but would be too slow for secure operation of small island systems like the AIPS.  
Here, it is relevant to note that the procurement of PFC services such as, for example, those procured by the GB TSO (see Fig.~\ref{fig:diffpfc}) have become cost-effective with relatively low prices (e.g., 1-5 £/MW/h).  

It is also worth observing that the frequency control, as well as the voltage control, remains intrinsically hierarchical but timescales are shifted of two orders of magnitude and `compressed' inside the converters.  These, in fact, include an inner control, an outer control and the actual frequency/voltage control, each of which is separated 
by an order of magnitude in the frequency domain to prevent undesirable couplings.  The need for such a hierarchy is for stability concerns.  However, the reason why conventional frequency control is assigned a specific set of timescales is strictly dependent on the dynamics of synchronous machines and their turbine governors.  With power electronic converters, the control, while remaining hierarchical, can be made faster.

\subsection{Different and Adaptive PFC Services}
\label{sec:pfc strategies}

\begin{figure}[htb]
  \begin{center}
    \resizebox{0.9\linewidth}{!}{\includegraphics{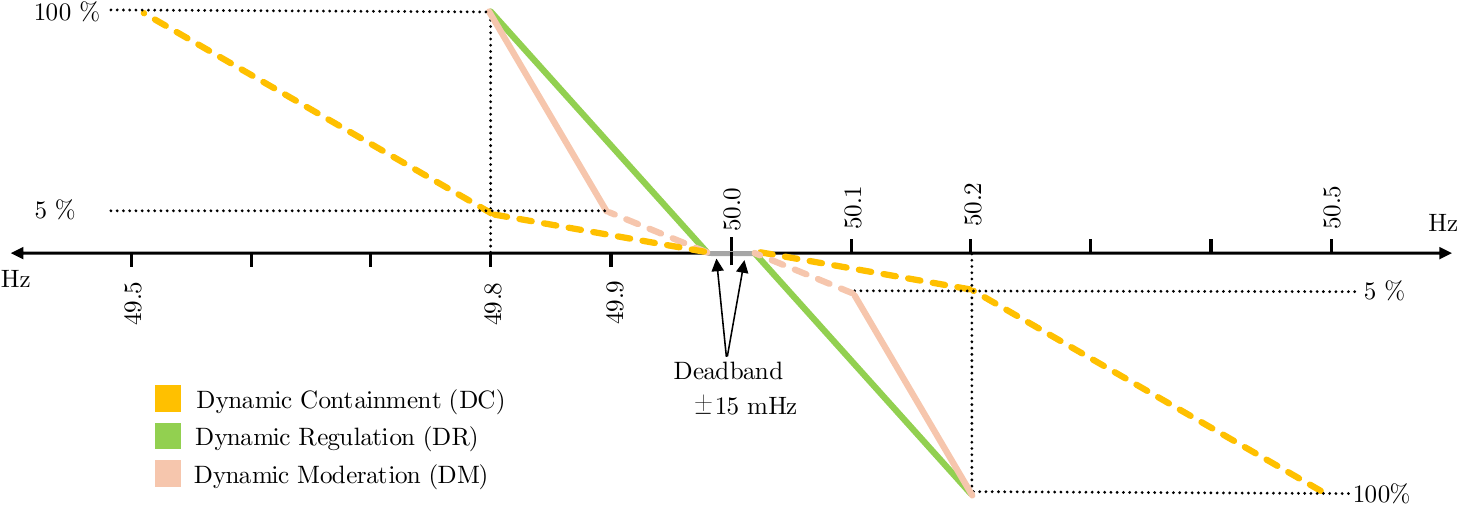}}
    \caption{New PFC reserve products introduced in GB (adapted from \cite{NESOpfr}).}
    \label{fig:diffpfc}
  \end{center}
  \vspace{-3mm}
\end{figure}

As discussed above, the proposed concept relies on strict mandatory PFC requirements and well-defined PFC ancillary services.  But in contrast to mandatory PFC that provides a defined response for both normal and abnormal conditions in terms of fixed parameters (e.g., deadband of $\pm$15 mHz, droop $\leq$ 5\% and response time $\leq$ 10s as is the case in AUS), TSOs may want, depending on system needs, to separate the tasks of dealing with normal and abnormal operating conditions and define/procure different PFC ancillary services.  
This is the case, for example, in GB where the TSO has defined three distinct PFC products namely dynamic containment (DC), dynamic moderation (DM) and dynamic regulation (DR) to deal with abnormal (DC) and normal (DR and DM) conditions, respectively.  These products are depicted in Fig.~\ref{fig:diffpfc}.  Note that similar to the AIPS GB does not implement an AGC.

As can be seen, DC with a response time of 1s is aimed at tackling post-fault operating conditions whereas DM (response time of 1s) and DR (response time of 10s) address pre-fault operating conditions in terms of keeping frequency within operational limits namely $\pm$200 mHz range.  In particular, observe that DM and DC apply adaptive droop/trajectories, that is, relatively little MW provision when frequency is within $\pm$100 mHz range and $\pm$200 mHz range, respectively.  

The Irish TSOs also apply adaptive PFC.  For instance, they can change the BESS PFC deadbands and droops/trajectories based on system conditions.  Similarly, the Irish TSOs also have the ability to change in real-time the deadbands of wind and solar plants, that is, $\pm$15 mHz or $\pm$200 mHz if frequency regulation and stability is challenging or not challenging, respectively.  As a matter of fact, the Irish TSOs have recently changed the PFC deadbands of wind and solar units to $\pm$15 mHz to address potential OF issues driven by lack of fault ride-through from demand facilities.  Therefore, in practice, narrow PFC deadbands are also used to deal with contingency conditions and not just to regulate the frequency under normal operating conditions.  In particular, having narrow deadbands on fast-acting resources such as BESS could significantly address frequency nadir/zenith issues as shown in \cite{kerci2025frequency}.   It can be concluded that a key advantage of today's and future power systems is the ability to implement (if required) adaptive PFC.  This was not possible in conventional power systems where it was hard, if not impossible, to change, for example, the droops or deadbands of synchronous generators in real-time.

\section{Economic Considerations}
\label{sec:market}
 
The proposed concept assumes the existence of a real-time energy market that co-optimizes the procurement of energy and PFC reserves (see Fig.~\ref{fig:newpfcservices}).  It is also prudent that such markets integrate balancing and congestion/stability management (i.e., network and stability constraints are explicitly modeled in the optimization process) to ensure overall system efficiency.  These markets reduce the gap between market outcomes and physical conditions or constraints of the power system.  For example, if the output of such markets reduce overall power imbalance, less frequency control will be needed and the less TSOs will need to intervene or correct operational issues.  

The closest real-world energy markets that reflect such characteristics are most of the North American real-time markets and that in AUS.  In particular, AUS uses a self-commitment central real-time dispatch market (no day-ahead market) that runs every 5 minutes and produces dispatch instructions for scheduled units.  Because of the short timescales these markets also help deal with contingencies, in particular, non-credible contingencies that push frequency outside defined operational limits.  For example, due to significant cascading events in AUS on 13/02/2024 at 13:08, AEMO lost approximately 2.7 GW of generation and 1 GW of load was shaken off in Victoria following the disturbance.  Redispatch of the real-time energy market occurred at 13:10 for the 13:15 trading interval and, thus, acted to successfully replace lost generation commencing within 5 minutes \cite{aemorecovery}.  Due to similar timescales, the dynamics of such markets overlap with the dynamics of AGC (see Table \ref{tab:different}).  For this reason,  one may argue that similar short cycle markets will effectively replace most, if not all, of the AGC objectives as they can be considered as some sort of market-based AGC or non-perfect-tracking AGC \cite{9361269}.  But observe that in contrast to AGC, real-time energy markets have an important advantage because only some resources participate under AGC while in general all resources above a certain MW size (e.g., $\geq$ 10 MW) must participate in energy markets.


Moreover, from a frequency control perspective we are of the view that market arrangements such as those in CE that strongly rely on unconstrained wholesale energy markets and market participants or balancing responsible parties (BRPs) for maintaining the balance between generation and demand (self-dispatch models) are not efficient.  For instance, despite BRPs being responsible financially for any imbalances they cause in the system, they are more focused on economic aspects rather than helping to keep the overall power balance which is the task of TSOs.  These market arrangements are also known to lead to significant deterministic frequency deviations \cite{CELICORTES20251029}.

Finally, another critical aspect of efficient and proper functioning energy markets are accurate demand and generation forecasts.  As a matter of fact, market participants and TSOs are heavily investing in advanced and reliable forecasting.  Therefore, it could be expected that in the future demand and generation forecasts, in particular short-term ones from a frequency perspective, will improve. But of course there could be instances where forecasts will be wrong for different reasons.   It is important that TSOs have the necessary mitigation measures in place to deal with such situations.  For instance, during the third quarter of 2025 AUS had one event where frequency was outside the normal band ($\pm$150 mHz) for more than five minutes without an
identified contingency event \cite{aemoerror}.  This event happened due to a significant self-forecasting error of around 1.8 GW of RES with a single
forecasting provider that affected several generators leading to frequency exceeding the 50.15 Hz operational limit.  Once the issue was detected by the provider, the self-forecasts for all relevant generators were suppressed resulting in the dispatch forecasts being sourced from AEMO’s internal
wind and solar forecasting systems, mitigating the issue.  Therefore, such rare events should be able to be primarily dealt with by market mechanisms, relevant TSO operational policies and then by abundant PFC available from both ancillary markets and mandatory requirements.

\section{Case Study}
\label{sec:case}

To complement the analysis under conditions not directly comparable in real cases, we perform simulations using Dome \cite{6672387} on a detailed dynamic model of the AIPS. This model, developed from publicly available data provided by the Irish TSOs, represents a realistic large-scale network comprising 1,479 buses, 1,851 transmission lines, 20 conventional synchronous generators, and 302 wind power plants.  Four operating scenarios are considered to evaluate the impact of controllers operating at different timescales:

\begin{itemize}
    \item Scenario 1: Conventional operation with AGC \cite{9361269}, with gain set to $k_0 = 10$.  AGC setpoints are issued every 2 s.
    \item Scenario 2: Same as Scenario 1, plus PFC-enabled wind generators using a deadband of $\pm 15$ mHz.
    \item Scenario 3: Same as Scenario 2 but without AGC.
    \item Scenario 4: Synchronous machines are replaced by GFM. $100\%$ IBR-based resources (GFM+Grid-following), without AGC.
    
\end{itemize}

To excite the system dynamics, a continuous active and reactive power perturbation is applied to the network loads.  This disturbance follows an Ornstein–Uhlenbeck stochastic process and has a total magnitude equivalent to $10\%$ of the system demand, distributed across the load buses.  On top of the continuous disturbances, complementary perturbations are included to capture different operative conditions of the system.  These are implemented as load ramps and stochastic jumps.  Ramps are modeled as 5\% variations of the load, lasting 5 minutes each, as follows: (i) positive ramps around 7:00 am to model morning load pickup; (ii) negative ramps around 12:00 pm to represent PV injection; and (iii) positive ramps around 6:00 pm to represent the evening peak.  Stochastic jumps are modeled as discussed in \cite{jumps} and emulate short-term intermittency of RES as sudden load variations of $2.5\%$.  The primary variables of interest are the deviation of the system center-of-inertia frequency, $\Delta \omega_{\mathrm{CoI}}$, and the percentage of time during which the frequency exceeds a given threshold of $\pm$200 mHz, over a total time of simulation of 24 h.

\begin{table}[h]
\centering
\caption{Frequency statistics of the AIPS under stochastic perturbations.}
\label{tab:ireland_stats}
\begin{tabular}{lcc}
\hline
Scenario & $\sigma_{\rm CoI}$ [mHz] & \% Time Out of Range \\
\hline
Conv + AGC & 177 & 12.25 \\
Conv + AGC + PFC & 116 & 8.64 \\
Conv + PFC & 118 & 9.12 \\
100\%  IBR & 86 & 3.82 \\
\hline
\end{tabular}
\end{table}

The results of Table \ref{tab:ireland_stats} reveal a progressive tightening of the frequency distribution as faster control mechanisms are introduced. Scenarios 1 and 4 represent the extreme cases, exhibiting the widest and narrowest frequency distributions, respectively. Most notably, Scenarios 2 and 3 exhibit nearly identical statistical performance, despite the absence of AGC in Scenario 3. This observation highlights the key physical insight that PFC provided by inverter-based wind generators efficiently operates on shorter timescales than the SFC.  As a result, PFC directly suppresses frequency excursions at their origin, significantly reducing the magnitude and persistence of deviations before slower secondary controllers become active. Consequently, the contribution of AGC becomes marginal when effective PFC is present, as the primary disturbance is already mitigated by fast electromagnetic and converter-driven control actions. 
These findings confirm that PFC from IBRs fundamentally reshapes the hierarchy of frequency control, shifting the dominant stabilization mechanism from slow centralized regulation toward fast, decentralized IBRs.

\section{Concluding Remarks}
\label{sec:conclu}

This paper proposes to `re-engineer' the frequency control of current and future converter-dominated power grids.  That is, this position paper rethinks the `old' hierarchal structure of frequency control that is based PFC, SFC, TCR and TEC and proposes a new one that is based on PFC and real-time (e.g., 5-minute) energy markets.  The main findings and contributions of the paper are summarized below.

\begin{itemize}
    \item While the existing frequency control structure has some pros such as being based on many years of operational experience it also shows several important drawbacks.  In particular, it is slow; difficult to adapt to new technologies; and does not recognize developments in markets.
    \item Mandatory stringent PFC requirements (e.g., with $\pm$15 mHz deadband), are critical in addressing several aspects of frequency such as improved frequency stability, recovery and long-term frequency quality.
    \item IBR-based resources such as BESS, HVDC units and RES are shown to provide excellent PFC.  In particular, despite PFC from RES introducing asymmetry it allows keeping frequency within a tight range.
    \item If PFC reaches some critical mass of MW/0.1 Hz, relative to typical MW contingency sizes, AGC becomes unnecessary.
    \item TEC, TCR and tie-line control are less and less relevant to frequency control.
    \item Size of power systems alone is not an important factor but rather width of the frequency deadbands, the response capabilities of the generation/demand resources as well as the level of MW noise/sizzle the system experiences relative to its size are equally important as well. 
    \item Based on the above measurement-based findings, it proposes a new frequency control management approach that relies on PFC and a real-time energy market.
    \item The viability of the proposed concept is supported by dynamic simulations on a realistic model of the AIPS.  It presents opportunities for reduced ancillary services costs and power system complexity.
    \item The proposed concept appears compatible with real-time electricity markets. 
\end{itemize}


Future work will focus on further assessments of the proposed concept including detailed dynamic simulations for 100\% IBR-based power systems, and real-world trials to gain confidence with stakeholders, in particular, TSOs and RAs.  


\bibliographystyle{IEEEtran}
\bibliography{references}

\end{document}